\documentclass[runningheads]{llncs}
\usepackage{graphicx}
\usepackage{lmodern}
\usepackage{exscale}
\usepackage{lineno}

\usepackage{scalerel}
\usepackage{tikz}
\usetikzlibrary{svg.path}

\usepackage{amsmath}
\usepackage{tabularx}
\usepackage[numbers,sort&compress]{natbib}

\usepackage{hyperref}

\usepackage{tikz,xcolor}

\definecolor{lime}{HTML}{A6CE39}
\DeclareRobustCommand{\orcidicon}{
	\begin{tikzpicture}
	\draw[lime, fill=lime] (0,0) 
	circle [radius=0.16] 
	node[white] {{\fontfamily{qag}\selectfont \tiny ID}};
	\draw[white, fill=white] (-0.0625,0.095) 
	circle [radius=0.007];
	\end{tikzpicture}
	\hspace{-2mm}
}

\foreach \x in {A, ..., Z}{\expandafter\xdef\csname orcid\x\endcsname{\noexpand\href{https://orcid.org/\csname orcidauthor\x\endcsname}
			{\noexpand\orcidicon}}
}

\begin{document}
\title{An Upper Bound on the Complexity of Tablut}
%
%
\author{Andrea Galassi\orcidA{}}
%
\authorrunning{A. Galassi}
%
\institute{Department of Computer Science and Engineering (DISI) \\ University of Bologna, Bologna, Italy \\
\email{a.galassi@unibo.it}\\
}
\maketitle              
\begin{abstract}
Tablut is a complete-knowledge, deterministic, and asymmetric board game, which has not been solved nor properly studied yet. In this work, its rules and characteristics are presented, then a study on its complexity is reported. An upper bound to its complexity is found eventually by dividing the state-space of the game into subspaces according to specific conditions. This upper bound is comparable to the one found for Draughts, therefore, it would seem that the open challenge of solving this game requires a considerable computational effort.

\keywords{
	Board game  
	\and game-playing
	\and game complexity
	\and Tafl games.
}
\end{abstract}
%
%
%


\section{Introduction}

Tablut is a strategy board game belonging to the family of Tafl games, a group of Celtic and Nordic asymmetric board games designed for two players, which share similar rules. Tafl games (sometimes called Hnefatafl games) may derive from the Roman game Ludus latrunculorum, and have evolved into many different variants of the original game, such as Tablut, Brandubh, Hnefatafl, and Tawlbwrdd.
The exact rules of these games are difficult to known, since little documentation has survived until the present days, and Tablut is probably the one for which most information is available.

Indeed, Carl Nilsson Linnaeus wrote notes about the game in the XVIII century, after observing S\'ami tribes playing it, even if he didn't know their language, so he could not be sure about the original rules. In the following centuries, other researchers~\cite{murray1,murray2,Helmfrid,ashton2010linnaeus,ucs1264} have analyzed, translated, and adjusted Linnaeus' notes, producing a set of rules which makes the game relatively balanced and hopefully similar to its original version.
Due to this scarce documentation, Tafl games have never been subject to proper analysis, and therefore we don't know their complexity, nor their solution to any degree.\footnote{
    According to~\cite{searching-solutions}, a game can be solved in several degrees: ultra-weakly solved (the game-theoretic value for the initial position is known), weakly solved (a strategy is known to obtain the game-theoretic value of the game), and strongly solved (for any state reachable from the initial position, a strategy to obtain the game-theoretic value of that state is known.)}
The purpose of this work is to provide an initial estimate of the complexity of Tablut, with the hope to lay the grounds for deeper studies.

In Section~\ref{sec:rules}, the rules and the properties of Tablut are presented. In Section~\ref{sec:ub} the game is analyzed so to compute an upper bound on its complexity. Section~\ref{sec:conclusion} concludes.

\section{Rules and Properties}
\label{sec:rules}
Many different variants of Tablut rules do exist. In this work, the rules of the game are described mostly following the work of Ashton~\cite{ashton2010linnaeus}.

\subsection{Terminology, Material, and Setup}
\begin{figure}[tbp]
\centering
\includegraphics[width=.9\linewidth]{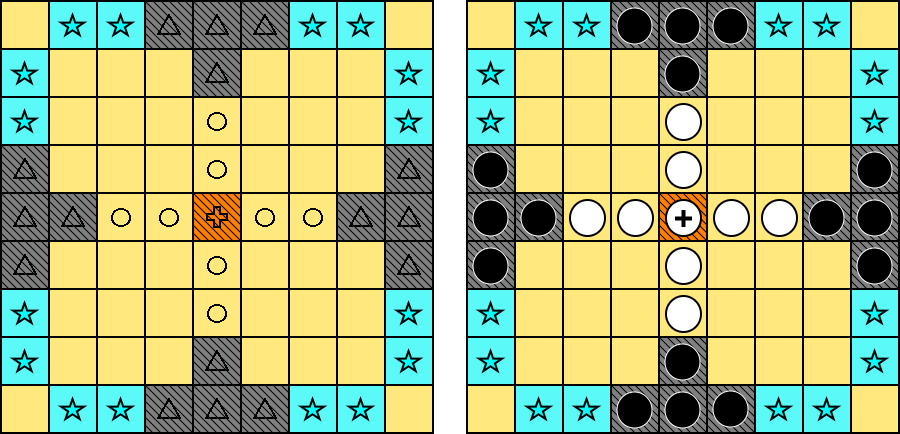}
\caption{The Tablut empty board (left) and the initial setup (right). The \emph{castle} cell is represented with the orange color and the cross symbol. The \emph{camp} cells are represented in grey with a triangle. The \emph{escape} cells are in blue, with a star. The white soldiers starting cells are marked with a circle. The \emph{king} checker is marked with a black cross symbol.
}
\label{fig:boards}
\end{figure}

The game is played by two players on a square board of $9\times9$ cells depicted in Figure~\ref{fig:boards}. The central cell is called the royal citadel, or castle, or throne. On each board side, there are 4 groups of 4 cells arranges with a t-shape that are called citadels or camps. For better comprehension, the former will be addressed as \textit{castle}, the latter as \textit{camps}, while the term \textit{citadel} will be used with the meaning ``either of them''.
Any non-camp and non-corner cell along the edge of the board will be addressed as \textit{escape}, the reason will be clear in the next Subsection.
Two cells are considered adjacent if and only if they are aligned horizontally or vertically and they share an edge. The term \textit{side} of a checker will be used to indicate any cell which is adjacent to the cell where the checker is placed.

One player moves the white checkers, which represent the defenders or Swedes, while the other moves the black checkers, which represent the attackers or Muscovites.
There are 16 black checkers and 9 white checkers.\footnote{
In this work, the terms \emph{checker}, \emph{stone}, and \emph{piece} will be used as synonyms.
}
One of the white checkers is the \textit{king} and it is marked. Any non-king piece will be addressed as \textit{soldier}. The checkers are placed as in Figure~\ref{fig:boards}, with the king in the castle, the black soldiers in the camps, and the remaining white soldiers aligned by 2 on each side of the king.

\subsection{Endgame}
The game ends when one of the following conditions is met:
\begin{enumerate}
\item The king reaches one of the escape cells. This results in the winning of the white player.
\item The king is captured. This results in the winning of the black player.
\item The game reaches the same state twice. This results in a draw.\footnote{
    This rule is added to avoid an endless sequence of repeated moves and to simplify the game with respect to \cite{ashton2010linnaeus}.
}
\item When a player has no possible moves. This results in the winning of the other player. This is an addition with respect to \cite{ashton2010linnaeus}.
\end{enumerate}

\subsection{Movement and Capture}
The two players alternate their turns, which consist of a single movement of a checker. 
The white player starts first. A checker can be moved along a single straight line, horizontally or vertically, by any number of cells. The movement must not pass over nor end into a cell that is occupied by another checker. The same holds for cells that are part of a citadel unless the checker starts its movement from a cell of the same citadel. This implies that the only checkers that may move inside a citadel are the black checkers, but only in their starting citadel and only if they have not ever left it.

To make a capture, a player must move one of its own pieces so as to surround an adversary piece. A checker is considered surrounded according to different criteria:
\begin{itemize}
\item When the king is in the castle, it is considered surrounded if there are enemy pieces on all four of its sides.
\item When the king is adjacent to the castle, it is considered surrounded if there are three enemy pieces on three of its side and the castle on the fourth.
\item When a soldier is adjacent to a citadel, or when the king is adjacent to a camp, it is considered surrounded if there is an enemy piece on the opposite side with respect to the citadel/camp. A soldier inside a camp cannot be captured.
\item In any other case, any piece is considered surrounded if there are two enemy pieces on two opposite sides of its cell so that the three pieces are aligned horizontally or vertically. 
\end{itemize}

Capture can happen only in an active way, which means that if a player moves their own piece so to make it surrounded, the piece is not captured. It is possible to capture multiple pieces (up to 3) with a single move if that move allows surrounding more than one piece.

\subsection{Properties}
At any moment of the game, the two players know everything about the state of the game. Given the state of the game, it is also known the consequence of any possible move, since there are no random components. Finally, the two players have different starting positions, checkers, and goals.
Therefore this is a complete-knowledge, deterministic, and asymmetric game.

The board is symmetric along 4 axes, so a single board configuration can have up to 7 symmetric configurations.

\section{Computing an Upper Bound for the State Space Size}
\label{sec:ub}
To compute the complexity of the game it is useful to divide the state-space into subspaces and compute their size separately. Firstly two subspaces are distinguished: the space without endgame states and the space with endgame states. Then, additional subspaces are considered according to specific conditions.
The symmetries of the problem will not be taken into account.

\subsection{State Space without Endgames}
It is useful to compute the complexity of the game without considering endgame positions because the computation of the final state of the game may not be necessary for some solvers. Indeed, the characteristics of the final move may be sufficient to determine the conclusion of the game and its winner, reducing the computational footprint. For example, a move that declares an escape cell as the destination for the king is an example of this.

Two possible endgames are not taken into account in this work: the endgame by draw and neither the endgame given by the impossibility for a player to move if he has still checkers on the board. The former is naturally included in the subspace without an endgame state. The latter is not investigated in this work.

It is possible to make the following considerations regarding any state that is not an endgame state:

\begin{itemize}

\item{The king has to be on the board, it can not be in a corner cell, and it can not be on escape cells, otherwise it would mean that one of the players has already won (respectively, the black or the white). Therefore, it can be in any of the $7\times7=49$ cells in the center of the board. Excluding the camps, there are only 45 possible cells, one of which is the castle.}
\item{There has to be at least one black soldier on the board. Otherwise, it would mean that the white player has won. Indeed, the black player could lose its last checker(s) only due to capture by the white player. The turn after that move, the black player would not be able to move any checker and therefore would lose.}
\item{The castle can host only the king.}
\end{itemize}

A naive initial upper bound can be computed considering the values that any cell can assume: 2 for the castle (with and without the king), 2 for the 16 camps cells (with and without the black soldier), 3 for the 20 edge cells (with a black soldier, a white soldier or neither), 4 for any other of the remaining 44 cells (with a black soldier, a white soldier, the king, empty). This would result in $10^{41}$ possible states as given by Equation~\ref{eq:states-naive}.

\begin{equation}
\label{eq:states-naive}
UB_{naive} = 2 \cdot 2^{16} \cdot 3^{20} \cdot 4^{44} \approx 10^{41}  
\end{equation}

To reduce the upper bound, it is possible to split the state space according to particular conditions.
It is possible indeed to differentiate two scenarios: when the king is in the castle and when it is not.

In the first scenario, each possible configuration has $1 \leq b \leq 16$ black pieces, $0 \leq w \leq 8$ white soldiers, and $e=80-w-b$ empty cells. It can therefore be modeled as a permutation with repetitions as in Equation~\ref{eq:p1}.

\begin{equation}
\label{eq:p1}
P_{80}^{(b,w,e)}=\frac{80!}{b! \cdot w! \cdot e!}
\end{equation}

The second scenario can be defined similarly, but the king would occupy one of the 44 cells (the castle is excluded). The remaining cells are therefore 79, so the number of states (for each combination of $b$, $w$, and $e$ value) is $44 \cdot P_{79}^{(b,w,e)}$.

An upper bound to the number of possible states without endgame positions is, therefore, $6\times10^{27}$, as given summing these two scenarios for any configuration of $b$, $w$, and $e$,  as in Equation~\ref{eq:states}.

\begin{equation}
\label{eq:states}
UB_{no\_end} = \displaystyle\sum_{b=1}^{16} \displaystyle\sum_{w=0}^{8} P_{80}^{(b,w,80-b-w)}  + 44 \cdot \displaystyle\sum_{b=1}^{16} \displaystyle\sum_{w=0}^{8} P_{79}^{(b,w,79-b-w)} \approx 6.1 \times 10^{27}
\end{equation} 

It is possible to consider one more characteristic: the white checkers cannot occupy a camp. Therefore, it is useful to treat camps and non-camps cells separately.
A new variable $c$ is now defined as the number of black checkers inside any camp. The number of black checkers outside the camp is then $b-c$. The number of possible configuration of black checkers inside the camps is now given by the permutation $P_{16}^{(c,16-c)}$, with $0 \leq c \leq b$.
The number of possible configurations on the non-camps cells can be computed as previously (considering 16 cells less).
According to this new consideration, an upper bound on the size of the two discussed scenarios is then computed as:

\begin{equation}
\label{eq:no-end-2-castle}
UB_{no\_end}^{castle} = \displaystyle\sum_{b=1}^{16} \displaystyle\sum_{w=0}^{8} \displaystyle\sum_{c=0}^{b} P_{16}^{(c,16-c)} \cdot P_{64}^{(b-c,w,64-b-w+c)}   \approx 3 \times 10^{25}
\end{equation} 

\begin{equation}
\label{eq:no-end-2-nocastle}
UB_{no\_end}^{no\_castle} = 44 \cdot \displaystyle\sum_{b=1}^{16} \displaystyle\sum_{w=0}^{8} \displaystyle\sum_{c=0}^{b} P_{16}^{(c,16-c)} \cdot P_{63}^{(b-c,w,63-b-w+c)}   \approx 9.2 \times 10^{26}
\end{equation} 

Their sum, (Equation~\ref{eq:no-end-2}) provides an upper bound $UB_{no\_end}' \approx 9.5 \times 10^{26}$

\begin{equation}
\label{eq:no-end-2}
UB_{no\_end}' = UB_{no\_end}^{castle} + UB_{no\_end}^{no\_castle} \approx 9.5 \times 10^{26}
\end{equation} 

Another fact that could be taken into account is that if three black pieces are in the ends of a camp, also the fourth black piece has to be in that camp, but this is will not be considered in this work.

\subsection{State Space of Endgames}
To make a proper comparison with other board games, an upper bound has to be computed considering also the endgame positions. This is done considering all the possible endgame scenarios computing an upper bound for each one.

Excluding the case when the king is captured, a first endgame scenario is obtained considering the states when all the black soldiers have been captured. As done previously, it is possible to divide the scenarios where the king is inside or outside the castle.

\begin{equation}
\label{eq:states-alpha}
UB_\alpha = \displaystyle\sum_{w=0}^{8} P_{64}^{(w,64-w)} +  44 \cdot P_{63}^{(w,63-w)}\approx 2.0 \times 10^{11}
\end{equation}

Another endgame scenario is given by the successful escape of the king to one of the 16 escape cells. The cell adjacent to the cell used to escape has to be empty, therefore only 62 cells can host black and white soldiers. In these cases, the lower bound on $b$ increases once again to 1, since without black soldiers the game would have already been finished.\footnote{It is possible for the king to make a capture in the same movement in which it reaches the escape, but this does not change the number of cases. On the opposite, this consideration would lower the number of states.} This contributes to the final upper bound with a term

\begin{equation}
\label{eq:states-beta}
UB_\beta = 16 \cdot \displaystyle\sum_{b=1}^{16} \displaystyle\sum_{w=0}^{8} \displaystyle\sum_{c=0}^{b} P_{16}^{(c,16-c)} \cdot P_{62}^{(b-c,w,62-b-w+c)}   \approx 2.3 \times 10^{26}
\end{equation}

The last endgame condition is given by the capture of the king, which could occur in many different cases. In the following scenarios, the variable $b$ will assume the meaning of the number of black soldiers on the board, that are not involved in the capture of the king.

\begin{enumerate}
\item When the king is inside of the castle, 4 black soldiers are necessary to capture it, therefore those checkers and those cells are determined. The number of cells which can host any other checker is therefore reduced from 64 to 60, while the highest possible value of $b$ is 12. The contribution of this case is:

\begin{equation}
\label{eq:states-gamma}
UB_\gamma = \displaystyle\sum_{b=0}^{12} \displaystyle\sum_{w=0}^{8} \displaystyle\sum_{c=0}^{b} P_{16}^{(c,16-c)} \cdot P_{60}^{(b-c,w,60-b-w+c)}   \approx 2.8 \times 10^{22}
\end{equation} 

\item When the king is adjacent to the castle (4 possible positions), 3 black soldiers are necessary to capture it. As for the previous case, the cells surrounding the king are for sure occupied by 3 black checkers. This reducing the number of cells to consider to 60 and the highest value of $b$ to 13. The upper bound is therefore computed as follows:

\begin{equation}
\label{eq:states-delta}
UB_\delta = 4 \cdot \displaystyle\sum_{b=0}^{13} \displaystyle\sum_{w=0}^{8} \displaystyle\sum_{c=0}^{b} P_{16}^{(c,16-c)} \cdot P_{60}^{(b-c,w,60-b-w+c)}   \approx 5.1 \times 10^{23}
\end{equation} 

\item When the king is adjacent to a camp (12 possible positions), 1 black soldiers is sufficient to capture it. In 8 case, the capturing checker can be in 2 positions, while in 4 it has to be in a specific position, therefore the possible scenarios are 20. The highest value of $b$ is 15 and the number of cells to consider are 62. The possible configurations for this scenario are:

\begin{equation}
\label{eq:states-epsilon}
UB_\epsilon = 20 \cdot \displaystyle\sum_{b=0}^{15} \displaystyle\sum_{w=0}^{8} \displaystyle\sum_{c=0}^{b} P_{16}^{(c,16-c)} \cdot P_{62}^{(b-c,w,62-b-w+c)}   \approx 8.0 \times 10^{25}
\end{equation} 

\item Finally, when the king is captured in any other position (28 possible cells), 2 black soldiers are necessary. For any position, there are 2 possible configurations of black soldiers, so the cases are 56. The highest number of black soldier on the board not involved in the capture is 14, and the cells to be taken into account are 61. So the possible configurations for this scenario are :

\begin{equation}
\label{eq:states-zeta}
UB_\zeta = 56 \cdot \displaystyle\sum_{b=0}^{14} \displaystyle\sum_{w=0}^{8} \displaystyle\sum_{c=0}^{b} P_{16}^{(c,16-c)} \cdot P_{61}^{(b-c,w,61-b-w+c)}   \approx 1.6 \times 10^{26}
\end{equation} 

\end{enumerate}

Taking all these scenarios into account as in Equation~\ref{eq:states-with-end}, the upper bound on the number of endgame states is $UB_{end} \approx 4.6 \times 10^{26}$.

\begin{equation}
\label{eq:states-with-end}
UB_{end} = UB_\alpha + UB_\beta + UB_\gamma + UB_\delta + UB_\epsilon + UB_\zeta \approx 4.6 \times 10^{26}
\end{equation} 

\subsection{Total State Space}

Summing the contribution of Equation~\ref{eq:no-end-2} and Equation~\ref{eq:states-with-end}, the final upper bound on the state space is $UB_{end} \approx 1.4 \times 10^{27}$:

\begin{equation}
\label{eq:states-final}
UB_{end} = UB_{no_end}' + UB_{end} \approx 1.4 \times 10^{27}
\end{equation} 

Taking into account possible endgames given by impossibility to move, or some other properties, would lead to an improvement of this estimation. For now, it is fair to assert that would be difficult to find even a weak solution for this game, since games with a similar state space are still unsolved, as illustrated in Table~\ref{tab:games}.

\begin{table}[htp]
 \caption{Comparison of upper bound on state-space complexity in different board games
 }
 
 \centering
 \begin{tabularx}{0.99\textwidth}{lccccccc}

 \noalign{\smallskip}
 & Tablut & \begin{tabular}{@{}c@{}}Nine Men's \\ Morris\end{tabular} &
 \begin{tabular}{@{}c@{}}English \\ Draughts\end{tabular} & \begin{tabular}{@{}c@{}}International \\ Draughts\end{tabular} & Othello & Chess  & Go \\
  \noalign{\smallskip}
 \hline
  \noalign{\smallskip}
$UB$ &  $1.4 \times 10^{27}$  & $3 \times 10^{11}$ &  $5 \times 10^{20}$ & $10^{30}$ & $10^{28} $ & $10^{43}$, $10^{50}$ & 2 $ \times 10^{170}$ \\
Solution  & No & Strong & Weak & No & No & No & No \\
Source &  & \cite{solving-NMM} & \cite{solving-draughts} & \cite{searching-solutions} & \cite{searching-solutions} & \cite{searching-solutions} & \cite{Go-states}  \\
\hline

 \end{tabularx}
 \label{tab:games}
\end{table}

\section{Conclusion and Discussion}
\label{sec:conclusion}
For the first time, an upper bound for the state-space complexity of the board game Tablut has been computed. The game seems to be comparable with the game of Draughts, therefore finding the strong solution of Tablut would probably require a great computational effort.
Nonetheless, many characteristics of the game have not been taken into account, therefore it is still possible to reduce this upper bound. Moreover, since no lower bound has been computed, its real complexity may be greatly inferior to what has been estimated in this work.

Due to the separation of the sub-spaces of the game, it is possible to know which are the scenarios that contribute the most to this upper bound. 
For what concerns the non-endgame subspace, they are the cases where the king is not in the castle. Among the end-game scenarios, they are the scenarios where the king escapes successfully.
Hopefully this initial investigation will encourage to compute a more accurate evaluation of the complexity of this game and to find a solution for it.

\bibliographystyle{splncs04}
\bibliography{biblio}

\end{document}